\documentclass{article}
\usepackage{spconf,amsmath,epsfig}
\usepackage{amsfonts,amssymb}
\usepackage{xcolor}
\usepackage{booktabs}
\usepackage{color}
\usepackage{soul}
\usepackage{fancyhdr}

\title{AETomo-Net: A Novel Deep Learning Network for Tomographic SAR Imaging Based on Multi-dimensional Features}
\name{Yu Ren, Xiaoling Zhang, Yunqiao Hu, Xu Zhan}
\address{University of Electronic Science and Technology of China}
\begin{document}
%
\maketitle
\begin{abstract}
Tomographic synthetic aperture radar (TomoSAR) imaging algorithms based on deep learning can effectively reduce computational costs. The idea of existing researches is to reconstruct the elevation for each range-azimuth cell in one-dimensional using a deep-unfolding network. However, since these methods are commonly sensitive to signal sparsity level, it usually leads to some drawbacks like continuous surface fractures, too many outliers, \textit{et al}. To address them, in this paper, a novel imaging network (AETomo-Net) based on multi-dimensional features is proposed. By adding a U-Net-like structure, AETomo-Net performs reconstruction by each azimuth-elevation slice and adds 2D features extraction and fusion capabilities to the original deep unrolling network. In this way, each azimuth-elevation slice can be reconstructed with richer features and the quality of the imaging results will be improved. Experiments show that the proposed method can effectively solve the above defects while ensuring imaging accuracy and computation speed compared with the traditional ISTA-based method and CV-LISTA.
\end{abstract}
\section{Introduction}
\label{sec:intro}

\begin{figure*}[htb]
\begin{minipage}[b]{1.0\linewidth}
    \centering
    \includegraphics{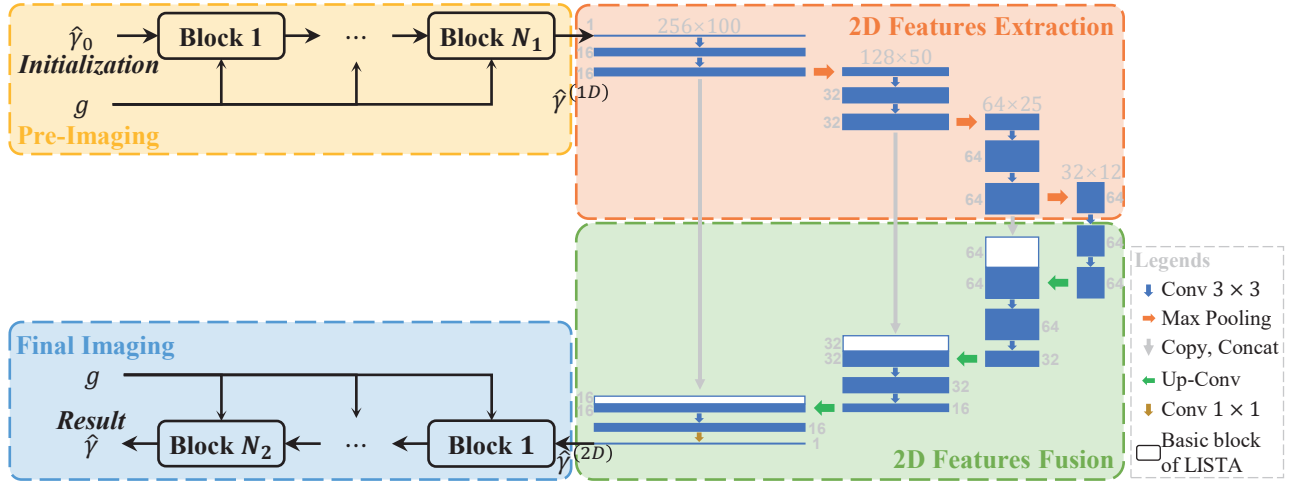}
\end{minipage}
    
    \caption{Structure of AETomo-Net. It consists of four parts. In this figure, each structure in a rounded rectangle is a part of AETomo-Net. The blue solid rectangles represent the size of feature maps. And other legends are marked in the lower right corner of the figure. In our experiments, we set $N_1 = 16$ and $N_2 = 32$.}
    
    \label{fg:diagram}
\end{figure*}

Tomographic synthetic aperture radar (TomoSAR) imaging, a technology that uses multiple angle observations to achieve the three-dimensional reconstruction of the imaging scenario, has been extensively studied in the past few decades. It has been widely used in earth remote sensing, resource exploration, and many other fields.

A typical class of methods for elevation reconstruction is the one based on compressed sensing (CS) theory. They complete the elevation reconstruction by solving a sparse linear inverse problem \cite{zhu2010-cs-tomo}. There have been some methods proposed based on it \cite{zhu2010-cs-tomo}, \cite{shi2018fast}, and they have achieved good performance. However, as they commonly do reconstruction iteratively, it will lead to an expensive computational cost \cite{cvlista}.

Recently, deep learning (DL) has been successfully applied in natural language processing, images classification, recommendation system, and many other fields. Some methods, like LISTA \cite{lista}, TISTA \cite{trainable-ista}, \textit{et al}, begin to use DL to solve the sparse linear inverse problem. Once training, the model can apply to all the reconstruction tasks under the same scenario without any other iterative calculations. Compared with the traditional CS method, it will greatly reduce the computational complexity.

Now, a few studies, like CV-LISTA \cite{cvlista} have used deep learning to solve the problem of tomography reconstruction. They regard elevation reconstruction as a one-dimensional process and use the LISTA-like method to reconstruct separately for each range-azimuth cell. Using these methods, the computational costs can be greatly reduced while maintaining good performance. However, these methods are sensitive to signal sparsity level \cite{block-sparse}. The results of them usually have some phenomenon like fractures of continuous planes, too many outliers, \textit{et al}. It seriously affects the quality of imaging results.


When using LISTA-like methods, because of the way they are performed, they only reconstruct based on one-dimensional information. But in the TomoSAR results, there is a certain correlation between adjacent pixels, that is, there are multi-dimensional features that we can use. If these multi-dimensional features are added to the TomoSAR imaging process, the reconstruction can be completed based on more abundant features. In this way, the imaging quality can be improved.

U-Net \cite{unet} is a CNN architecture that makes full use of 2D features. Its contracting path completes the successive extraction of 2D features of different depths, and the expansive path fully integrates and utilizes 2D features of different depths.


Inspired by it, we propose our network called AETomo-Net to perform TomoSAR imaging using multi-dimensional features. Firstly we use a LISTA-like architecture to do pre-imaging for each range-azimuth cell. Then we add a fully convolutional structure to extract 2D features of azimuth-elevation slices. After that, the feature maps of different depths are fully fused. Finally, we use the other LISTA-like structure to perform the final imaging. Experiments show that the results of AETomo-Net have better continuity in the structure of the continuous plane and have fewer outliers compared with the traditional ISTA-based method and CV-LISTA. It indicates that the addition of CNN structure integrates 2D features of azimuth-elevation slices effectively into the imaging process.



\section{Methodology}
\label{sec:methodology}

According to the previous research \cite{zhu2010-cs-tomo}, observing the imaging scenario from multi baselines is equivalent to a process of sparse sampling, which can be expressed as
\begin{equation}
    g = R \gamma + n ,
\end{equation}
where, $g \in \mathbb{C}^M$ is the observation data, $R \in \mathbb{C}^{M \times N}$ is the measurement matrix, $\gamma \in \mathbb{C}^N$ represents the scattering characteristics of the observation scenario, $n$ is the noise, and $M \ll N$. The purpose of TomoSAR is to reconstruct $\gamma$ from the observation data $g$.

CV-LISTA \cite{cvlista} solve it by mapping an iteration of the traditional ISTA to a basic block expressed in equation (\ref{eq:listablock}) and solve it in learning way. It is a 1D process which do reconstruction for each range-azimuth cell respectively.
\begin{equation}
    \hat{\gamma_k} = \mathrm{soft}_{\theta^k}\left(W_1 g + W_2 \hat{\gamma}_{k-1}\right) . \label{eq:listablock}
\end{equation}

However, CV-LISTA is sensitive to sparsity level. So using it for TomoSAR can cause some problems like continuous surface breakage, too many outliers, etc.





\begin{figure*}[htb]
\begin{minipage}[b]{0.15\linewidth}
    \centering
    \centerline{\epsfig{figure=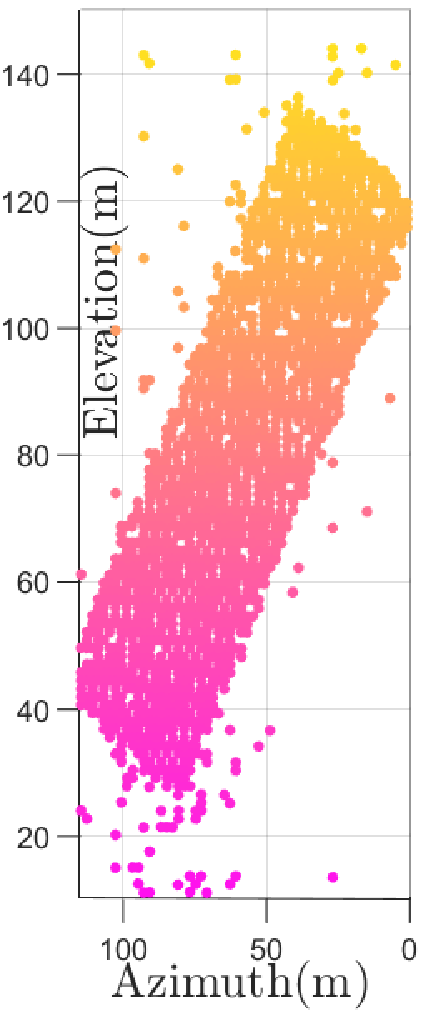, width=2.8cm}}
    \centerline{(a)}
\end{minipage}
\begin{minipage}[b]{0.15\linewidth}
    \centering
    \centerline{\epsfig{figure=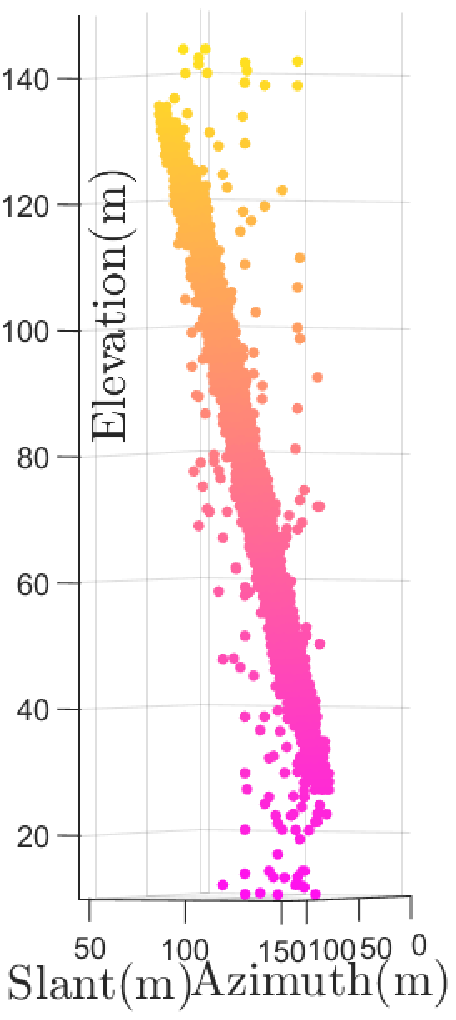, width=2.8cm}}
    \centerline{(b)}
\end{minipage}
\begin{minipage}[b]{0.15\linewidth}
    \centering
    \centerline{\epsfig{figure=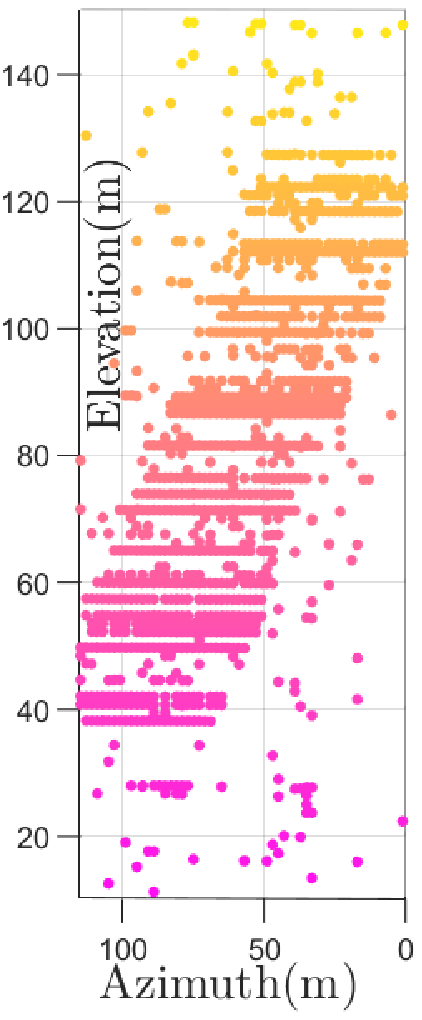, width=2.8cm}}
    \centerline{(c)}
\end{minipage}
\begin{minipage}[b]{0.15\linewidth}
    \centering
    \centerline{\epsfig{figure=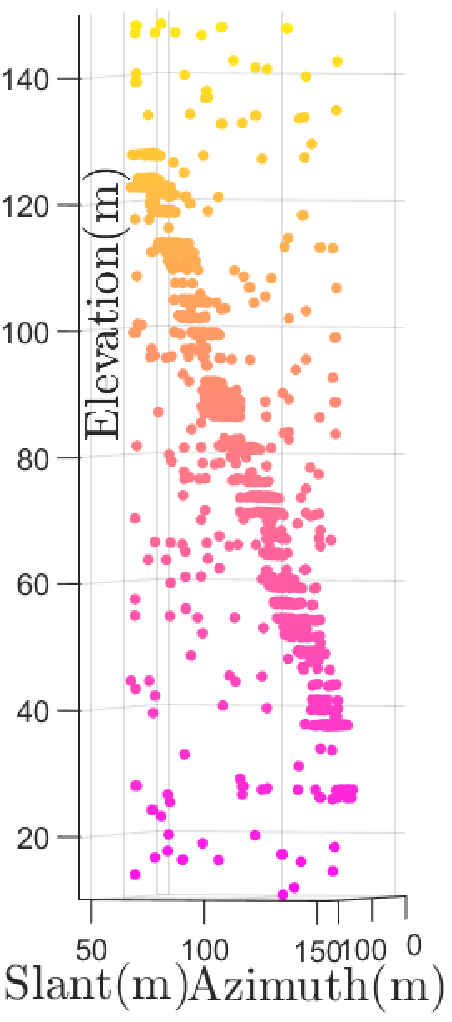, width=2.75cm}}
    \centerline{(d)}
\end{minipage}
\begin{minipage}[b]{0.15\linewidth}
    \centering
    \centerline{\epsfig{figure=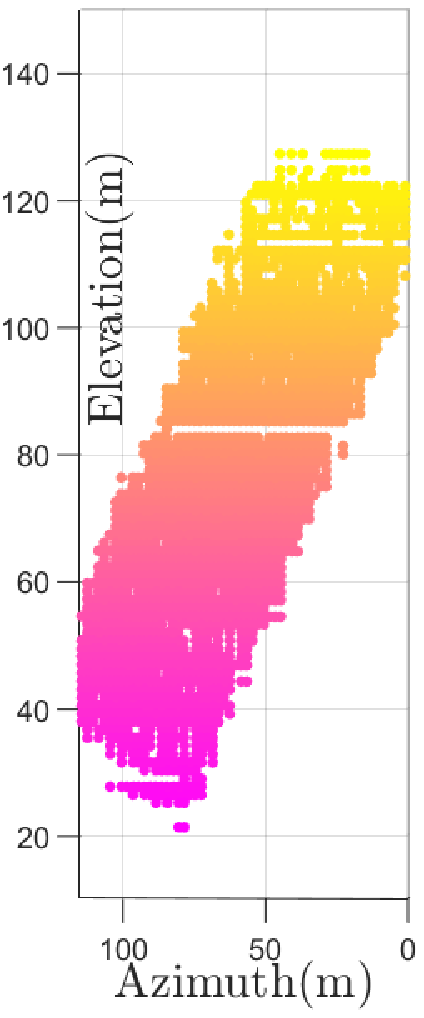, width=2.8cm}}
    \centerline{(e)}
\end{minipage}
\begin{minipage}[b]{0.15\linewidth}
    \centering
    \centerline{\epsfig{figure=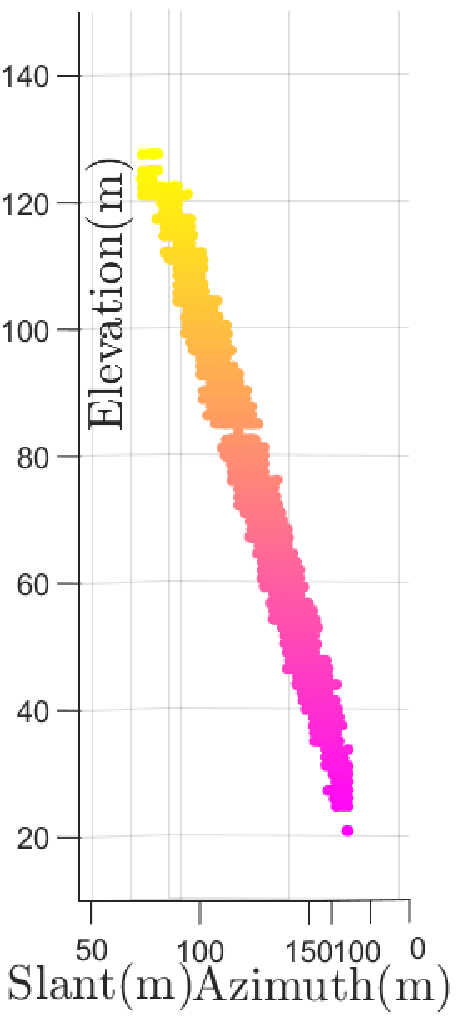, width=2.8cm}}
    \centerline{(f)}
\end{minipage}
\begin{minipage}[b]{.06\linewidth}
    \centering
    \centerline{\epsfig{figure=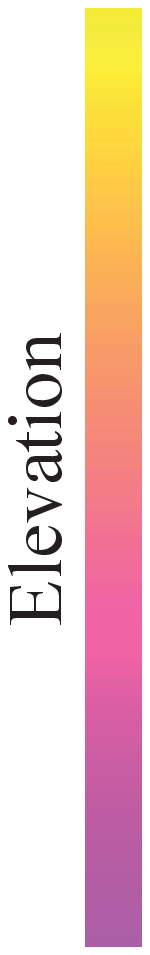, width = 1cm}}
   
\end{minipage}
    \caption{
    Demonstration of imaging results of a long oblique plane. (a) Front view of imaging result using (F)ISTA; (b) Left view of imaging result using (F)ISTA; (c) Front view of imaging result using CV-LISTA; (d) Left view of imaging result using CV-LISTA; (e) Front view of imaging result using AETomo-Net; (f) Left view of imaging result using AETomo-Net.}
    \label{fg:result}
\end{figure*}


This problem can be solved by adding abundant multi-dimensional features. In this paper, we propose a novel network called AETomo-Net to achieve that. Our main idea is imaging combined with multi-dimensional features. The architecture of it is illustrated in Fig. \ref{fg:diagram}.

In general, AETomo-Net consists of four parts, \textit{i.e.} pre-imaging, 2D features extraction, 2D features fusion and final imaging. By adding a fully convolutional structure to the deep-unfolding structure, we extract 2D features of azimuth-elevation slices and fuse them into 1D imaging data. In this way, we integrate 2D information into the process of the TomoSAR imaging algorithm, which is originally a 1D process, to improve its performance. The details of each part will be introduced as follows.

\subsection{Pre-Imaging}


Inspired by CV-LISTA, we design an RNN-like structure with $N_1$ blocks to do pre-imaging for each range-azimuth cell of the current processing azimuth-elevation slice.

In this part, the observation data is processed along one dimension. After performing through this structure, the TomoSAR observation data is initially focused along the elevation direction.

\subsection{2D Features Extraction}

The second part is composed of a series of convolution layers. It is used to extract 2D features of azimuth-elevation slices.

This part consists of three blocks. Each block uses two convolution layers with $3 \times 3$ kernel, each followed by a rectified linear unit (ReLU), and one $2 \times 2$ max pooling operation with stride $2$ to do downsampling. After each downsampling, the size of one single feature map is divided by $4$ and the number of feature maps is multiplied by $2$. Each time a block is processed, deeper 2D salient features will be extracted once. At the end of this stage, we will obtain a feature map containing the deep 2D features extracted from the current processing azimuth-elevation slice.

\subsection{2D Features Fusion}

We use a series of transposed convolution operations to finish 2D feature fusion. It is also composed of three blocks.

Before the start of each block, the size of the feature map is multiplied by $4$ using a $2 \times 2$ transposed convolution. Then merge it with the symmetric feature maps of the 2D feature extraction part and halve its number twice by $3 \times 3$ convolution layers. After multiple steps of up-sampling, merging, and twice convolutions, the 2D features of TomoSAR data at different depths are plenty fused.

\subsection{Final Imaging}

At the end of AETomo-Net, considering that the added 2D features will have redundancy, we use the other RNN-like structure with $N_2$ blocks to perform final imaging.


Similar to the first part, we use the other deep-unfolding structure to perform a 1D process to complete the final imaging of the data fused with 2D features. After this stage, we can get the final result of TomoSAR imaging.

\section{Experiments}

\subsection{Dataset}

AETomo-Net is trained on our simulated data. We firstly convert 3D models to point clouds using PCL 1.12.0. Then we simulated the observation data of different baselines in a similar way to RaySAR \cite{raysar-application}.

In our experiments, fourteen 3D models of architecture were used to finish this simulation task. We used $24$ uniform distributed baselines in the range of $-200 \mathrm{m}$ to $200 \mathrm{m}$ and set the height of the reference baseline to $5.0456 \times 10^5 \mathrm{m}$. The scenario center oblique distance $R_0$ was set to $6.1434 \times 10^5 \mathrm{m}$ and the radar incidence angle of the reference baseline was set to $34.78^\circ$. And, the SAR system we simulated worked in X-band, so we set the wavelength to $0.031 \mathrm{m}$.

\subsection{Data Loading}

To effectively use 2D features, we redesigned an approach of data loading. While loading data, the data of an azimuth-elevation slice is loaded instead of random data of multiple range-azimuth cells. In the first and fourth parts of AETomo-Net, like the previous algorithms, it is still a 1D process. But in the second and third parts, it is a 2D process. In this way, we can utilize AETomo-Net to efficiently acquire 2D features for TomoSAR imaging.

\subsection{Loss Function Design}

In the training of AETomo-Net, we need to consider three constraints to ensure that each part works well, which are, (1) the performance of pre-imaging, (2) the performance of features extraction and fusion, and (3) the final imaging effect.

Given a training data $\{g_i, \gamma_i^*\}$, we will get the pre-imaging result $\hat{\gamma}_i^{(1D)}$, the output data fused with 2D features $\hat{\gamma}_i^{(2D)}$ and the final imaging result $\hat{\gamma}_i$ in the process of AETomo-Net. For better imaging quality, we expect that these three results to be as close to the ground truth as possible. Meanwhile, we need to constrain the sparsity of the final imaging result $\hat{\gamma}_i$. Therefore, the loss function of AETomo-Net is designed as follows.

\begin{equation}
    \mathcal{L} = \mathcal{L}_{1D} + \alpha \mathcal{L}_{2D} + \beta \mathcal{L}_{im} ,
\end{equation}
with,
$$
\left\{
    \begin{array}{lr}
        \mathcal{L}_{1D} = \frac{1}{N_s} \sum_{i=1}^{N_s} \| \hat{\gamma}_i^{(1D)} - \gamma_i^* \|^2_2 & \\
        \mathcal{L}_{2D} = \frac{1}{N_s} \sum_{i=1}^{N_s} \| \hat{\gamma}_i^{(2D)} - \gamma_i^* \|^2_2 & \\
        \mathcal{L}_{im} = \frac{1}{N_s} \sum_{i=1}^{N_s} \left[ \| \hat{\gamma}_i - \gamma_i^* \|^2_2 + \lambda \| \hat{\gamma}_i \|_1 \right]
    \end{array} ,
\right.
$$
where $\alpha$, $\beta$ are weight terms to adjust the importance of each part in the training process, $\lambda$ is a trade off term to balance the error and sparsity of results and $N_s$ is the number of range-azimuth cells in a single azimuth-elevation slice. In our experiments, we set $\alpha = 0.6$, $\beta = 2.2$, $\lambda = 0.05$ and $N_s = 100$.

\section{Results}

Fig. \ref{fg:result} shows the demonstration of imaging results using (F)ISTA, CV-LISTA, and AETomo-Net respectively. Note that we use a long oblique plane to show the imaging quality. The plane is not parallel to any axes, which is more in line with the real situation.

Apparently, there are less fracture in Fig. \ref{fg:result}(e)(f) than Fig. \ref{fg:result}(c)(d). The reconstruction result using AETomo-Net is more continuous. And there are fewer outliers in the result of AETomo-Net compared with (F)ISTA and CV-LISTA. The accuracy of AETomo-Net is also higher.

Then we used three different metrics to evaluate the performance of (F)ISTA, CV-LISTA, and AETomo-Net according to the point cloud extracted from the TomoSAR imaging results. The computational speed of three different algorithms was also measured using the same personal computer. Table \ref{tab:compare} shows the comparison result.

\begin{table}[!htbp]
    \centering
    \caption{Comparison of TomoSAR imaging using the traditional (F)ISTA, CV-LISTA and AETomo-Net, where accuracy and completeness are defined by $A_{dist}$ and $C_{dist}$ in \cite{performance-measurement} to measure the performance of point cloud reconstruction.}
    \begin{tabular}{cccc}
    \toprule
                     & (F)ISTA & CV-LISTA & AETomo-Net \\
    \midrule
        Accuracy     & 3.7649  &  5.0303  & \textbf{2.2876} \\
        Completeness & \textbf{0.8137}  &  1.6685  & 0.9346 \\
        Outliers (\%)& 9.3867\% &  11.3518\% & \textbf{1.6615\%} \\
        Times        & hours  & \textbf{$<1$ min} & \textbf{$<1$ min} \\
    \bottomrule
    \end{tabular}
    \label{tab:compare}
\end{table}

As can be seen from Table \ref{tab:compare}, the results of AETomo-Net and (F)ISTA are much better than that of CV-LISTA in terms of completeness and accuracy. And the result of AETomo-Net is more accurate and has fewer outliers while having almost the same completeness as (F)ISTA. Meanwhile, the deep learning methods have a very huge reduction in computing time compared with the traditional (F)ISTA.




\section{Conclusion}

In this paper, we propose AETomo-Net to improve the performance of TomoSAR imaging by adding multi-dimensional features. We add a 2D feature extraction part and a 2D feature fusion part composed entirely of convolutional layers to the original imaging method based on deep-unfolding architecture, so that the network can use more abundant features for imaging. Experiments show that the result of AETomo-Net is more continuous and has fewer outliers compared with (F)ISTA and CV-LISTA. And using deep learning methods can greatly reduce calculating time. It shows that combining multi-dimensional features can improve the imaging quality of TomoSAR effectively. The utilization of multi-dimensional features and deep learning has great potential for TomoSAR high-precision imaging.




\bibliographystyle{IEEEbib}
\bibliography{main}

\end{document}